\documentclass{article}

\usepackage{amsfonts}
\usepackage{amssymb}
\usepackage{hyperref}
\usepackage[nosumlimits,nointlimits,nonamelimits]{amsmath}
\usepackage{amsmath}
\usepackage{graphicx}
\usepackage{caption}
\usepackage{subcaption}

\usepackage{relsize}

\usepackage{appendix}

\newtheorem{theorem}{Theorem}[section]
\newtheorem{proposition}[theorem]{Proposition}
\newtheorem{definition}[theorem]{Definition}
\newtheorem{lemma}[theorem]{Lemma}

\newenvironment{proof}[1][Proof]{\noindent \textbf{#1.} }{\  \rule{0.5em}{0.5em}}

\usepackage{color, soul}

\begin{document}

\title{Team Decision Problems with Classical and Quantum Signals\thanks{We are grateful to Samson Abramsky, Lucy Brandenburger, Jerome Busemeyer, Adan Cabello, Eugenia Cerda, Ignacio Esponda, Artem Kaznatcheev, Elliot Lipnowski, Hong Luo, Thomas Philippon, Gus Stuart, Natalya Vinokurova, Noson Yanosfsky, Sasha Zolley, and audiences at the CUNY Graduate Center, NYU, Yale, the Workshop on Correlated Information Change, University of Amsterdam, November 2014, and the ``Quantum Physics Meets TARK" workshop, University of Groningen, July 2011, for extremely valuable comments, and to the NYU Stern School of Business, J.P. Valles, and the HHL -- Leipzig Graduate School of Management for financial support.}}

\author{Adam Brandenburger\thanks{Stern School of Business, Polytechnic School of Engineering, Institute for the Interdisciplinary Study of Decision Making, Center for Data Science, New York University, New York, NY 10012, U.S.A., \href{mailto:adam.brandenburger@stern.nyu.edu}{adam.brandenburger@stern.nyu.edu}, \href{http:adambrandenburger.com}{adambrandenburger.com}}
\and
Pierfrancesco La Mura\thanks{HHL -- Leipzig Graduate School of Management, 04109 Leipzig, Germany, \href{mailto:plamura@hhl.de}{plamura@hhl.de}}}

\date{Version 12/11/14\\
\medskip
}
\maketitle

\begin{abstract}
We study team decision problems where communication is not possible, but coordination among team members can be realized via signals in a shared environment.  We consider a variety of decision problems that differ in what team members know about one another's actions and knowledge.  For each type of decision problem, we investigate how different assumptions on the available signals affect team performance.  Specifically, we consider the cases of perfectly correlated, i.i.d., and exchangeable classical signals, as well as the case of quantum signals.  We find that, whereas in perfect-recall trees (Kuhn \cite[1950]{kuhn-50}, \cite[1953]{kuhn-53}) no type of signal improves performance, in imperfect-recall trees quantum signals may bring an improvement.  Isbell \cite[1957]{isbell-57} proved that in non-Kuhn trees, classical i.i.d.~signals may improve performance.  We show that further improvement may be possible by use of classical exchangeable or quantum signals.  We include an example of the effect of quantum signals in the context of high-frequency trading.
\end{abstract}

\section{Introduction}
\thispagestyle{empty}
A team is a group of agents unified by common goals.  Characteristic of team problems is that members of a team have access to different information depending on their local environments.  Communication of this information among team members may or may not be possible, depending on economic and physical constraints.  An example of the latter arise in high-frequency trading (see Pagnotta and Philippon \cite[2012]{pagnotta-philippon-12} for a survey), where messaging across widely dispersed members of a team would be too slow to be useful.  In this paper, we study scenarios where direct communication is indeed unavailable, but team members can use a \textbf{shared global environment} to achieve highly effective coordination.  We undertake a systematic examination of how the informational properties of the environment interact with the informational structure of a decision problem to bring about changes in performance.

In the absence of communication, team problems become formally equivalent to one-agent decision problems with memory limitations.  This equivalence was noted by Marschak and Radner (\cite[1972]{marschak-radner-72}) in their pioneering work on teams.  We shall refer to these scenarios as \textbf{team decision problems}.  Figure 1 is a simple example.

\begin{figure}[here]
\quad\quad\quad\quad\quad\quad\quad\quad\quad\quad\quad\quad\quad\quad\quad\quad\quad\,
   \includegraphics[width=1.9in]{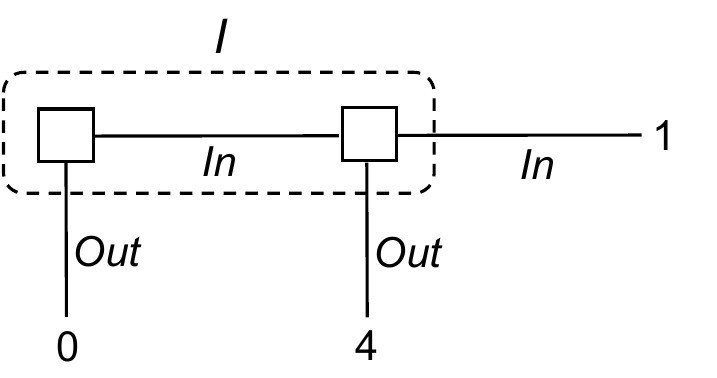}
      \begin{center}
   Figure 1
   \end{center}
\end{figure}

In the scenario accompanying this tree, there is a task that requires the completion of two different steps in sequence.  If the two steps are completed in the correct order, the payoff is $4$.  If the second step is undertaken before the completion of the first, the payoff is $0$.  If the first step is mistakenly repeated, the payoff is $1$.  There are two people assigned to the task and each has to act without knowing if the other has already completed the first step.  The tree of Figure 1 captures this scenario.  In particular, the two square nodes belong to the two team members, but the nodes are enclosed in an information set $I$ to capture the idea that they do not know whether they are acting first or second.\footnote{In the tradition of Savage \cite[1954, p.13]{savage-54} and his famous example of making an omelette, we could think of our task as flipping a frittata precisely once.  The task is to be performed by two distracted cooks, so that neither knows if the other has yet completed the task.  Of course, one flip is best (payoff of $4$), two flips leads to an overdone dish (payoff of $1$), and no flip leaves the eggs undercooked (payoff of $0$).}

As a one-agent problem, this same scenario has been extensively discussed as the Absent-Minded Driver's Problem (Piccione and Rubinstein \cite[1997]{piccione-rubinstein-97}).  While the focus in that literature was on conceptual aspects of this scenario, our interest is more `engineering-like.'  Specifically, we shall investigate how well a team can perform tasks such as the above one, as a function of the assumptions made on its shared environment.

A concrete example of how environmental information can affect performance in a decision problem of the type in Figure 1 was offered by Isbell \cite[1957]{isbell-57}.  He showed that if the players have access to \textbf{i.i.d.~signals} (payoff-irrelevant chance moves), then they can do better relative to no signals.  There are other possibilities.  Members of a team might have access to \textbf{exchangeable} (not necessarily i.i.d.) signals.  Will this make a difference --- in particular, will it allow still better performance?

Another possibility is that the physical make-up of the environment matters.  In other areas of information theory, it is well-established that access to \textbf{quantum} rather than \textbf{classical information resources} has profound consequences for various tasks.  One main distinction between classical and quantum signals is that they arise at different physical scales.  Classical signals are encoded in the macroscopic state of some physical system --- for example, in an electrical current or light beam (or even in smoke signals $\ldots$).  By contrast, quantum signals are encoded in the microscopic state of a system --- for example, in the spin of an electron or of a photon.  Most importantly, quantum signals can exhibit patterns of behavior that are impossible with any choice of classical signals.  In particular, quantum signals may be not only correlated but even \textbf{entangled}, where this term refers to exotic correlations that cannot arise in the classical case.\footnote{The paper is self-contained in that we will describe later the (little) physics the reader will need.  This said, Susskind \cite[2012]{susskind12} is an excellent introduction to the phenomenon of quantum entanglement.}

While quantum signals permeate any physical environment, their controlled use as information resources has only recently become possible and implementable.  One case in which quantum techniques has already entered the practical arena is quantum cryptography (Qiu \cite[2014]{qiu-14}), where the security of communication is protected by the very laws of Nature; by contrast, analogous classical schemes do not offer similar guarantees.  In computer science, there are important quantum algorithms that have been proved superior to classical algorithms (Deutsch-Jozsa \cite[1992]{deutsch-jozsa92}, Grover \cite[1996]{grover96}, Shor \cite[1997]{shor97}).  Could it be that in the area of decision making, as in cryptography and computing, the availability of quantum resources might lead to improved performance over what is possible in a classical environment?  We will identify conditions under which this is indeed the case.  This may not be of theoretical interest alone.  We will come back later to the example of high-frequency trading, where access to quantum resources could have practical significance.

Team problems with classical signals have been studied by Lehrer, Rosenberg, and Shmaya \cite[2010]{lehrer-rosenberg-shmaya10}.  Their focus is on signals which are informative about the underlying (`physical') state, while our interest is in signals as coordinating devices when there is a fixed information structure concerning the underlying state.  La Mura \cite[2005]{lamura05} provides an example of a team problem where quantum signals yield an improvement over classical signals.  (We make use of this example later.)  Kargin \cite[2008]{kargin08} provides a necessary condition for quantum signals to yield no improvement in a specific family of team problems.

\section{Results}
Kuhn \cite[1950]{kuhn-50}, \cite[1953]{kuhn-53} introduced into decision theory the fundamental distinction between \textbf{perfect} and \textbf{imperfect recall}.  Isbell \cite[1957]{isbell-57} extended this classification further to include other trees with limited recall, which do not belong to the family of Kuhn trees.  Those include decision problems in which, as in the example of Figure 1, an information set may contain nodes which are met in sequence.  We will call these \textbf{Isbell trees}.  This classification is equally important in team decision problems, where it refers to the availability or unavailability of information about what other team members do or know.  This three-way taxonomy of decision problems --- perfect-recall Kuhn, imperfect-recall Kuhn, Isbell --- is the one we will use.
\begin{figure}[here]
\quad\quad\quad\quad\quad\quad\quad\quad\quad\quad\quad\quad\quad
   \includegraphics[width=2.5in]{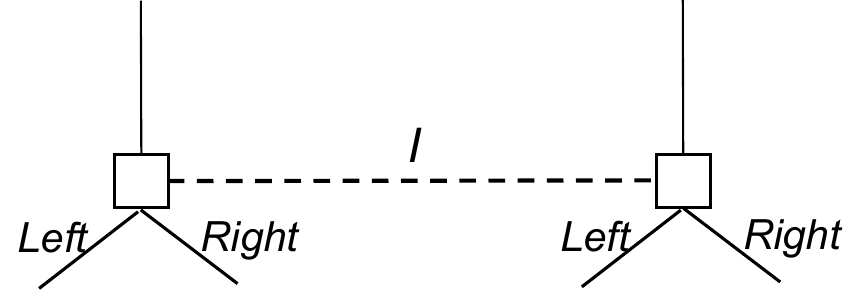}
      \begin{center}
   Figure 2
   \end{center}
\end{figure}

We now add a framework for talking about the different kinds of signals to which members of a team might have access.  In the simplest case, there is one signal per information set.  But this is restrictive and does not fit well with cases where the different nodes in a given information set could be widely separated from one another in space or time.  In such cases, it may be more appropriate to think of distinct but perfectly correlated signals operating at different nodes within the same information set.  In fact, other assumptions on signals are possible that still preserve the indistinguishability of nodes in an information set.  In particular, the signals could be \textbf{i.i.d.}, but, more generally, we can require them to be \textbf{exchangeable}.\footnote{Two random variables are exchangeable if their joint distribution is invariant under permutation (Billingsley \cite[1995, p.473]{billingsley95}.}  Figure 2 depicts an information set in some tree.  Figure 3a is the simple case where one coin is tossed at this information set and the choices can be pegged on the outcome of the toss.  Figure 3b depicts two coins, one per node, where the coins are exchangeable (which includes the case they are i.i.d.).

\begin{figure}[here]
\quad\quad\quad\quad\quad\quad\quad\quad\quad\quad\quad
   \includegraphics[width=3.8in]{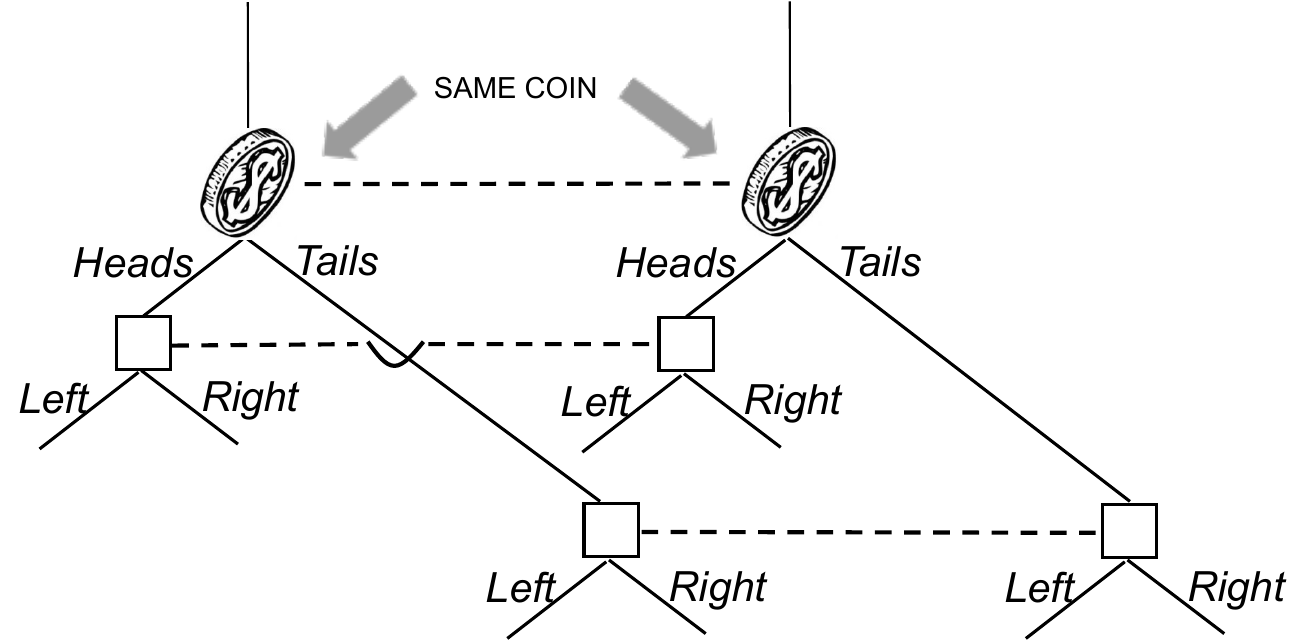}
      \begin{center}
   Figure 3a
   \end{center}
\end{figure}
\begin{figure}[here]
\quad\quad\quad\quad\quad\quad\quad\quad\quad\quad\quad
   \includegraphics[width=3.8in]{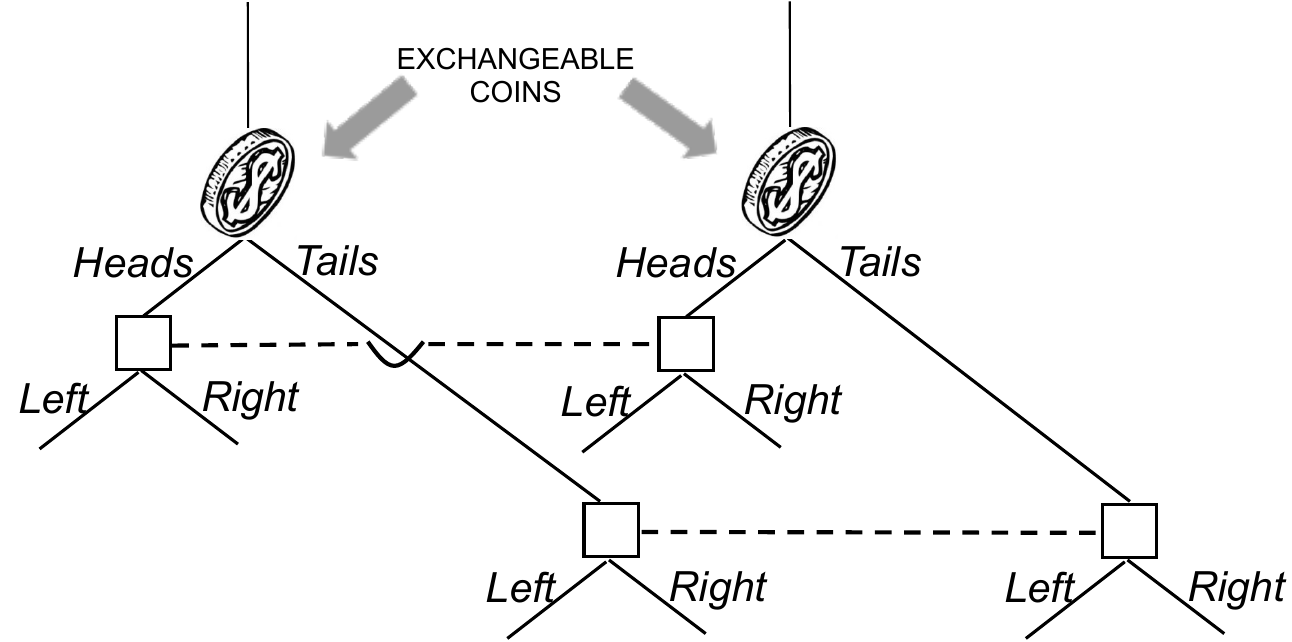}
      \begin{center}
   Figure 3b
   \end{center}
\end{figure}

Within information sets, there are some clear considerations of indistinguishability.  We also need to consider what are the appropriate conditions to impose on signals across information sets.  We will want to know how these conditions, too, affect the potential performance in a task.  To uncover these capabilities, it becomes important to specify the physical embodiment of the signals that are available.  In particular, what correlations across signals are possible fundamentally depends on whether the signal carrier obeys classical or quantum physical laws.

\begin{figure}[here]
   \includegraphics[width=5.5in]{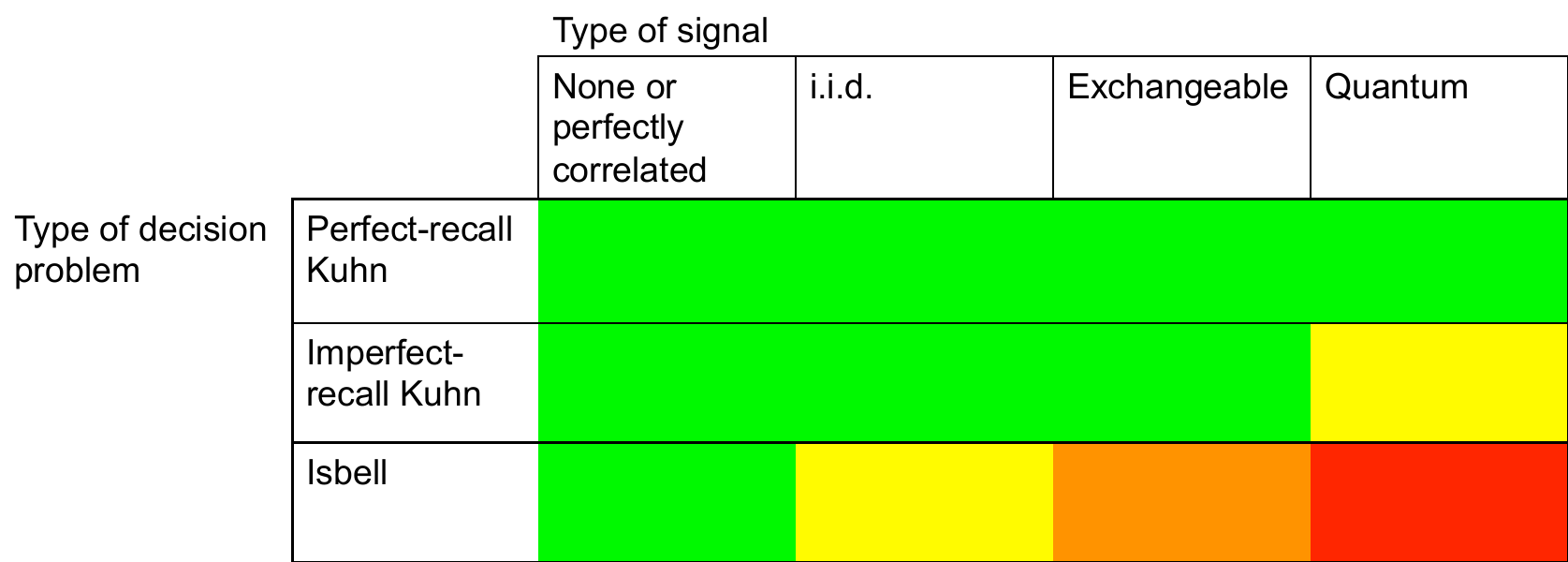}
   \begin{center}
   Table 1
   \end{center}
\end{figure}

Table 1 shows, for each type of decision problem we consider, the effect on team performance of different assumptions about the type of signals available.  We denote in green the baseline performance which can be achieved in all types of problem with no signals.  Along a given row, higher performance is indicated by moving from green to orange to yellow to red (as in a heat map).  Our results can be summarized as follows.  For perfect-recall Kuhn trees, no type of signal brings any improvement over the baseline.  For imperfect-recall Kuhn trees, no classical signal type (perfectly correlated, i.i.d., or exchangeable) can improve over the baseline, but quantum signals may do so.  For Isbell trees, classical i.i.d.~signals improve over the baseline, exchangeable signs improve further, and quantum signals still further.

We see from the table that in situations where communication among members of a team would be helpful but is unavailable, signals can act as substitutes, at least in part.  Another implication of our results is that decision making is not invariant to the physical embodiment of the decision environment.  In particular, we see that access to quantum signals may yield improvements over any choice of classical signals.  We will make last point more concrete via an example later in the paper.

\section{Signal Structures}
Figure 4 depicts a team decision problem which begins with a chance move.  This is represented by a circular node belonging to Nature, where the numbers in parentheses give the probabilities of Nature's move.  One team member, Ann, moves at information set $I_1$, and the other member, Bob, moves at $I_2$.  When Ann moves, she knows that Nature chose \textit{left}.   But, when Bob moves, he does not know whether Nature chose \textit{right}, or Nature chose \textit{left} and then Ann chose \textit{In}.  Thus, Ann may have information --- that Nature chose \textit{left} and she chose \textit{In} --- that Bob does not get.  In terms of our three-way taxonomy, the team problem is a Kuhn tree with imperfect recall.  The reader should refer to Definitions A.1 (Kuhn tree) and A.4 (perfect recall) in Appendix A to check this last statement.

\begin{figure}[here]
\quad\quad\quad\quad\quad\quad\quad\quad\quad\quad\quad\quad
   \includegraphics[width=3.0in]{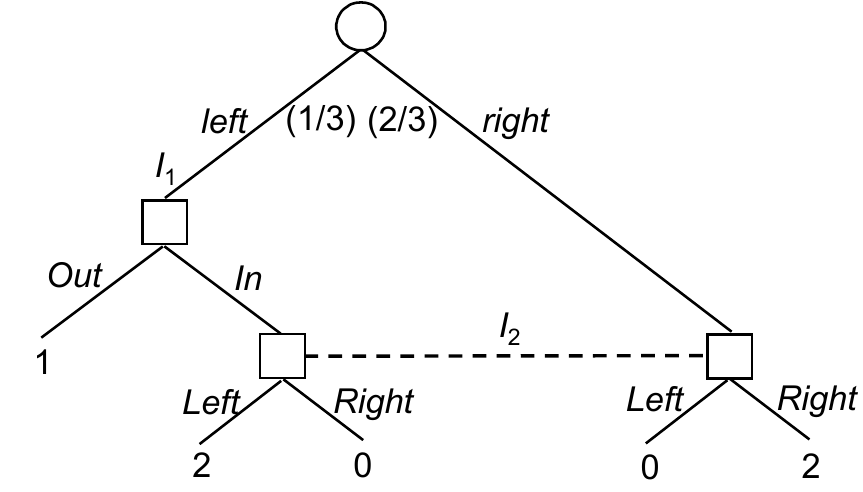}
   \begin{center}
   Figure 4
   \end{center}
\end{figure}

The team's expected payoffs are:~$2/3$ from the pair of strategies (\textit{In}, \textit{Left}), $4/3$ from (\textit{In}, \textit{Right}), $1/3$ from (\textit{Out}, \textit{Left}), and $5/3$ from \textit{Out}, \textit{Right}).  The team's highest expected payoff is therefore $5/3$.

Figure 5 adds two signals, in the form of coin tosses, to Figure 4.  There is one coin toss at information set $I_1$, with outcomes \textit{Heads}$_1$ and  \textit{Tails}$_1$, and another coin toss at $I_2$, with outcomes \textit{Heads}$_2$ and  \textit{Tails}$_2$.  Ann can make her choice at $I_1$ contingent on the outcome of the coin toss at $I_1$.  Likewise, Bob can make his choice at $I_2$ contingent on the outcome of the coin toss at $I_2$.

It will be convenient to describe the probability structure of the coin tosses in the following way.  Each possible path through a tree crosses certain information sets of the team members in a certain order.  A \textbf{signal structure} associates to each sequence of information sets that arises in this fashion, a probability measure on the product space of the associated signals.  In Figure 5, the sequences of information sets that can arise are $I_1$, $I_2$, and $I_1I_2$.  Figure 6 gives the general form of the associated probability measures.  Here $\alpha$ through $\theta$ are numbers between $0$ and $1$ satisfying $\alpha + \beta = 1$, $\gamma + \delta = 1$, and $\epsilon + \zeta + \eta + \theta = 1$.

\begin{figure}[here]
\quad\quad
   \includegraphics[width=5.5in]{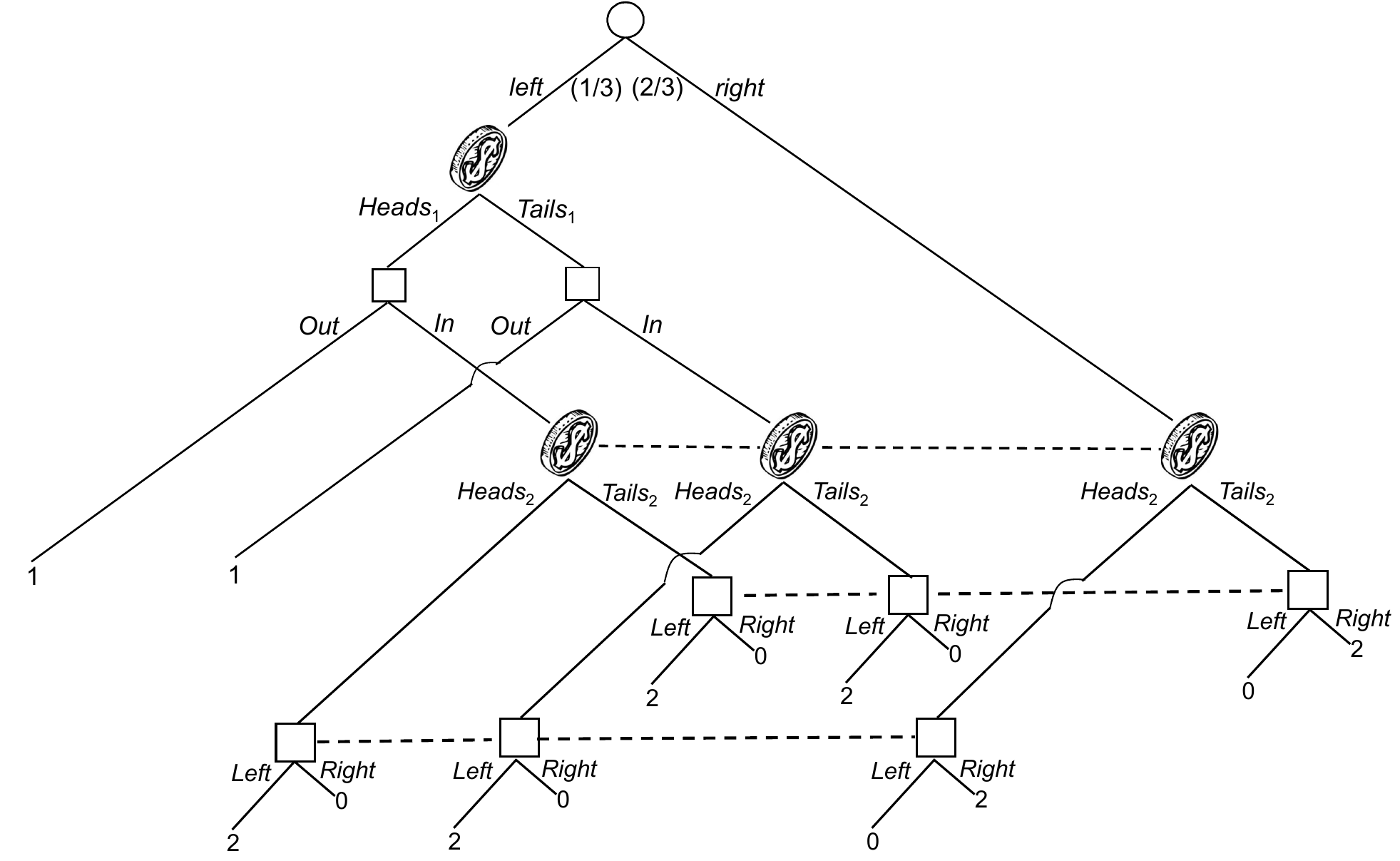}
   \begin{center}
   Figure 5
   \end{center}
\end{figure}

\begin{figure}[here]
\quad\quad\quad\quad\quad\quad\quad\quad\quad\quad\quad\,\,
   \includegraphics[width=2.5in]{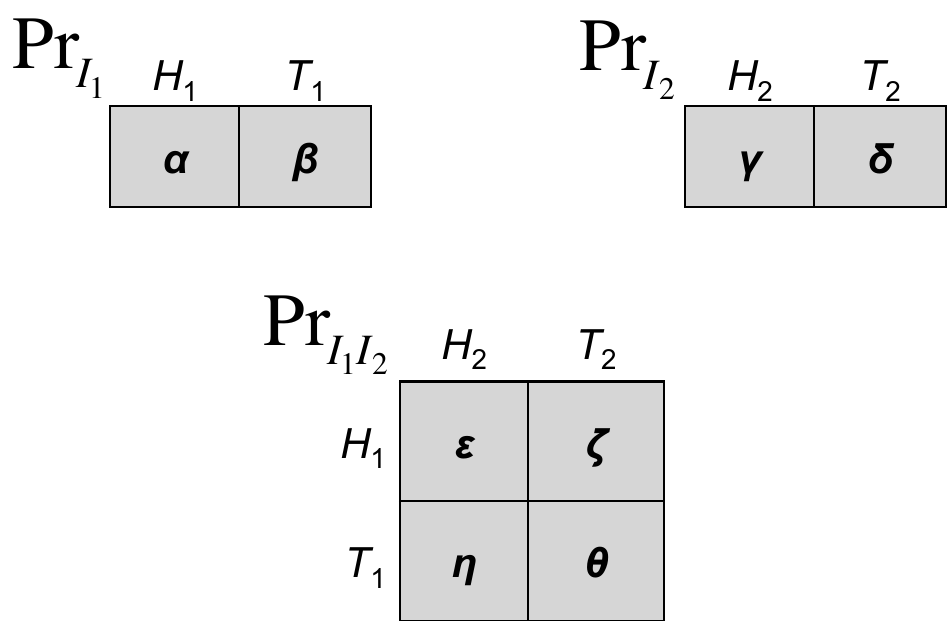}
   \begin{center}
   Figure 6
   \end{center}
\end{figure}

Consider the following strategies for the team.  At her information set, Ann chooses \textit{Out} if her coin comes up \textit{Heads}$_1$, and \textit{In} if her coin comes up \textit{Tails}$_1$.  At his information set, Bob chooses \textit{Left} if his coin comes up \textit{Heads}$_2$, and \textit{Right} if his coin comes up \textit{Tails}$_2$.  The team's expected payoff is then:
\begin{equation*}
1/3 \times (\alpha \times 1 + \eta \times 2 + \theta \times 0) + 2/3 \times (\gamma \times 0 + \delta \times 2).
\end{equation*}
If $\alpha = 0$, $\delta = 1$ (so that $\gamma = 0$), and $\eta = 1$ (so that $\theta = 0$), then the team gets an expected payoff of $2$, which is greater than the best possible ($5/3$) without signals.

On a closer look, it is apparent that the information structure of Figure 6 when $\alpha = 0$ and $\delta = \eta = 1$ is conceptually unsatisfactory.  The second coin comes up Heads ($H_2$) almost surely if it is tossed after the first coin has been tossed, but comes up Tails ($T_2$) almost surely if it is tossed when the first coin has not been tossed.  If we allow this amount of information `flow' between the two coins, it is not surprising that we can engineer an improvement over the case of no coins.  Our interest is in situations where such direct communication is impossible.  What is needed, then, is a condition that rules out such information flow or communication, without going so far as to rule out all correlation across signals.  The next section presents a suitable condition.

\section{Indistinguishability}
We want a condition that applies across information sets and that is in line with the basic requirement that nodes within an information set are indistinguishable.  Suppose, for a moment, that the outcomes of a signal at one information set could reveal, probabilistically at least, which other signals are activated elsewhere in the tree.  Then, simply by observation of the signal, a team member at a given information set might learn something --- probabilistically, say --- about which node in the information set was reached.  (This was the case in the example of Figures 5 and 6.  If Bob observes $H_2$, then he knows almost surely he is at the middle node of $I_2$.  If he observes $T_2$, he knows almost surely he is at the right-hand node of $I_2$.)  Here is the formal condition to rule out this possibility:

\begin{quote}
\textbf{Indistinguishability Condition}  Consider two sequences of information sets and the two associated signal probability measures.  The marginals of these two measures --- with respect to common subsequences --- must agree.
\end{quote}

Let's see how this condition works in the signal structure of Figure 6.  Looking at the two sequences $I_1$ and $I_1I_2$, with common subsequence $I_1$, we see that the condition is that the probabilities of $H_1$ must be equal: $\alpha = \epsilon + \zeta$.  Likewise, looking at the two sequences $I_2$ and $I_1I_2$, with common subsequence $I_2$, we see that the condition is that the probabilities of $H_2$ must be equal: $\gamma = \epsilon + \eta$.  Of course, the first (resp.~second) condition implies that the probabilities of $T_1$ (resp.~$T_2$) are also equal.

Indistinguishability rules out the choice of parameters $\alpha = 0$ and $\delta = \eta = 1$ we had before in Figure 6.  (This choice contradicts $\gamma = \epsilon + \eta$.)  We can go further.  Indistinguishability reduces the five free parameters in Figure 6 to three, which we will take to be $\epsilon$, $\zeta$, and $\eta$ (we will still write $\theta = 1 - \epsilon - \zeta - \eta$).  The expected payoff under the previous strategy can then be written as:
\begin{equation*}
\epsilon \times (1/3 \times 1 + 2/3 \times 0) + \zeta \times (1/3 \times 1 + 2/3 \times 2) + \eta \times (1/3 \times 2 + 2/3 \times 0) + \theta \times (1/3 \times 0 + 2/3 \times 2).
\end{equation*}
Since this is a convex combination of expected payoffs to the team in the tree without signals, we see that no improvement in the team's (maximum) expected payoff is now possible under signals.  The same is easily seen to be true for any other strategy for the tree of Figure 5.

\section{Classicality}
The example of the previous section might suggest that, provided the indistinguishability condition is satisfied, the addition of signals to a (Kuhn) decision problem can never result in an increase in the team's maximum expected payoff.  Table 1 tells us that this is false once quantum signals are allowed.  We will see an example of this phenomenon in the next section.  Before that, we will establish the correct baseline for signals to yield no improvement in Kuhn trees.

Fix a Kuhn tree, let $I_1, I_2, \ldots$ be the information sets for the DM and $\Omega_{I_1}, \Omega_{I_2}, \ldots$ be associated finite signal sets which we add.\footnote{If there is no signal at an information set, the signal space is a singleton.  For simplicity, we restrict attention to finite signal sets.}  Write $\Omega = \Omega_{I_1} \times \Omega_{I_2} \times \cdots$.

\begin{quote}
\textbf{Classicality Condition}  There is a probability measure $\mu$ on $\Omega$ such that for each subsequence $I_{i_1} I_{i_2}\cdots I_{i_K}$ of information sets that arises in the tree, the associated probability measure is given by:
\begin{equation*}
{\rm{Pr}}_{I_{i_1}I_{i_2}\cdots I_{i_K}} = {\rm{marg}}_{\Omega_{I_{i_1}} \times \Omega_{I_{i_2}} \times \cdots \times \Omega_{I_{i_K}}} \mu.
\end{equation*}
\end{quote}

\noindent Note that this condition is well-defined since, in a Kuhn tree, each path through the tree crosses a given information set at most once.  Classicality says that there is a joint probability space $(\Omega, \mu)$ from which a given signal structure, of the kind we explored in the previous section, can be derived.  It is immediate from the properties of marginals that:

\begin{proposition}
Classicality implies indistinguishability.
\end{proposition}

Next, let $M_{I_1}, M_{I_2}, \ldots$ be the sets of moves at the information sets $I_1, I_2, \ldots$ respectively, and write $M = M_{I_1} \times M_{I_2} \times \cdots$.

\begin{proposition}
Fix a Kuhn tree.  The highest expected payoff a team can achieve with signals satisfying classicality is the same as that without signals.
\end{proposition}

\begin{proof}
A strategy profile for the team in the underlying tree is an element $m \in M$.  A strategy profile for the team in the extended tree with signals is a tuple of maps $f_{I_1} : \Omega_{I_1} \rightarrow M_{I_1}, f_{I_2} : \Omega_{I_2}\rightarrow M_{I_2}, \ldots$.  Write $f = f_{I_1} \times f_{I_2} \times \cdots$.  Also, write $\pi(m)$ for the expected payoff to the team in the underlying tree, when it chooses strategy profile $m$ and we average over Nature.  Then, the expected payoff to the team in the extended tree, when it chooses strategy profile $f$ (and we again average over Nature), is $\sum_{m \in M}(\mu\circ f^{-1})(m)\times\pi(m)$.  That is, in the tree with signals, the expected payoff to any particular strategy profile is a convex combination of expected payoffs to strategy profile in the underlying tree.

This argument applies when there is one signal per information set.  Since in a Kuhn tree, every path from the root to a terminal node passes through a given information set at most once, it immediately extends to the case of multiple signals per information set, whether these signals are perfectly correlated, i.i.d., or exchangeable.
\end{proof}
\medskip

Throughout, we assume independence between Nature (in the underlying tree) and signals.  Independence seems like the right assumption for our purpose.  We do not want signals to give the team information it never had.  When independence is violated, Proposition 5.1 may fail.  Appendix B provides an example of such a case.

In decision theory (and game theory), one normally takes for granted the existence of a joint probability space that yields whatever signals one has in mind.  This makes sense in the classical physical world where every physical mechanism can be associated with an appropriate joint probability space.  But it may fail in the quantum realm.  It turns out that existence or non-existence of this joint space actually defines the classical-quantum divide (Fine \cite[1982]{fine82}, Abramsky and Brandenburger \cite[2011]{abramsky-brandenburger11}).  This is the reason behind the naming of our classicality condition.

\section{Quantum Improvement}
We now show:

\begin{proposition}
There is a Kuhn tree (with imperfect recall) in which the team can achieve a higher expected payoff with quantum signals than with any classical signals.
\end{proposition}

\begin{figure}[here]
\quad\quad\quad\quad\quad\quad\quad\,
   \includegraphics[width=4.0in]{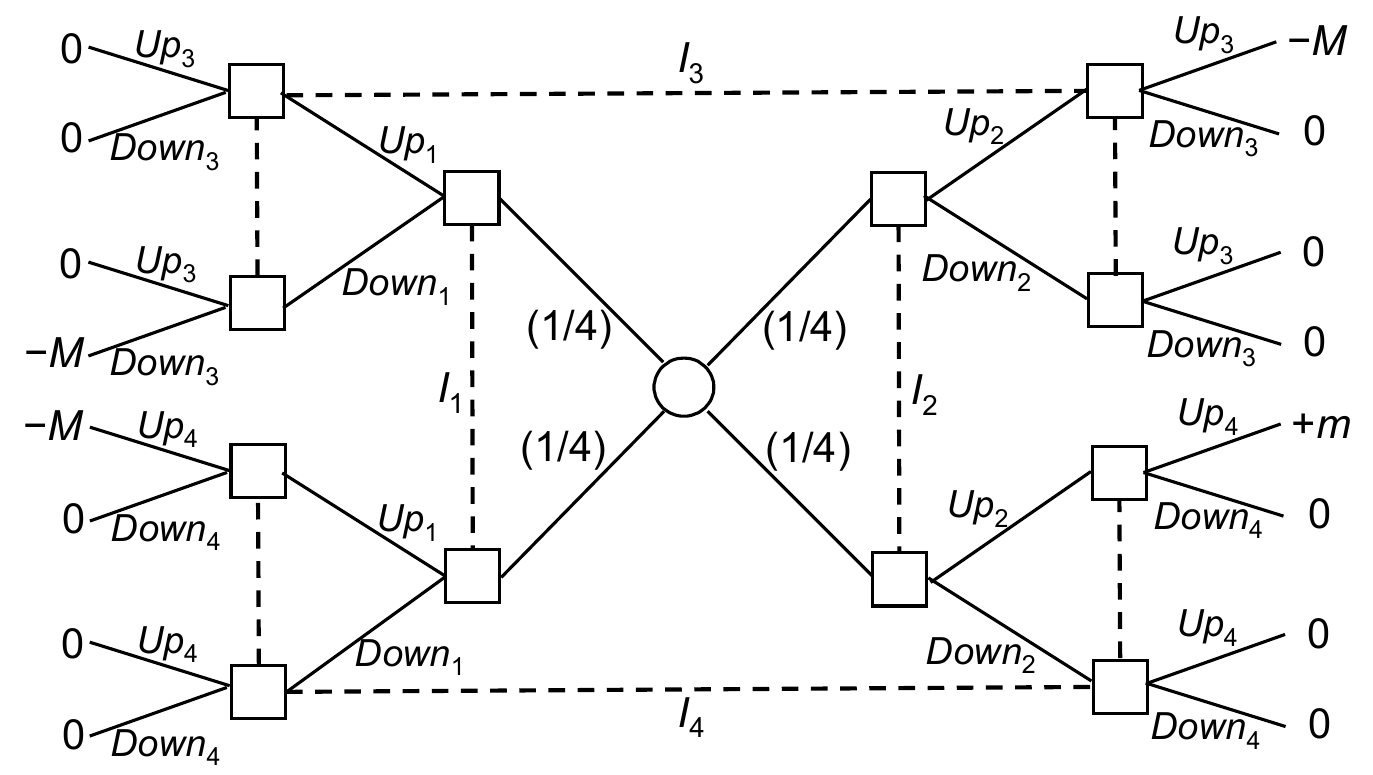}
    \begin{center}
   Figure 7
   \end{center}
\end{figure}

The decision problem of Figure 7 will suffice to establish this claim.  It represents a situation in which two team members are imperfectly informed of Nature's initial move and must coordinate their actions.  We assume that the payoffs satisfy $0 < m < M$.  We first show that the team's expected payoff with classical signals is at most $0$.  To see this, start without signals.  Observe that the only way for the team to get the $+m$ payoff (with positive probability) is if it chooses \textit{Up}$_2$ at information $I_2$ and \textit{Up}$_4$ at information set $I_4$.  But then, to avoid the $-M$ payoff on the right side of the tree, it must choose \textit{Down}$_3$ at $I_3$.  Then, to avoid the upper $-M$ payoff on the left side, it must choose \textit{Up}$_1$ at $I_1$.  Then, to avoid the lower $-M$ payoff on the left side, it must choose \textit{Down}$_4$ at $I_4$, not \textit{Up}$_4$ as we supposed.  It follows that the $+m$ payoff cannot arise unless at least one $-M$ payoff also arises.  Moreover, it will arise with the same probability.  Since $M > m$, we have shown that the team's expected payoff is at most $0$.  Now use Proposition 5.2 to conclude that the team's highest possible expected payoff in any extended tree with classical signals is also $0$.

Next, consider the signal structure of Figure 8.  Here, $\Phi = 2/(1 + \sqrt5)$ and is the inverse of the Golden Ratio.  One can check that our indistinguishability condition is satisfied (use the fact that $\Phi^2 + \Phi = 1$).  Now consider the following strategy profile for the team in the extended tree:~(i) at $I_1$, choose \textit{Up}$_1$ after $H_1$ and \textit{Down}$_1$ after $T_1$; (ii) at $I_2$, choose \textit{Up}$_2$ after $H_2$ and \textit{Down}$_2$ after $T_2$; (iii) at $I_3$, choose \textit{Up}$_3$ after $H_3$ and \textit{Down}$_3$ after $T_3$; (iv) at $I_4$, choose \textit{Up}$_4$ after $H_4$ and \textit{Down}$_4$ after $T_4$.  The team's expected payoff from this strategy profile is $1/4 \times \Phi^5 \times m > 0$.  An improvement in the team's highest expected payoff is achieved.

\begin{figure}[here]
\quad\quad\quad\quad\quad\quad\quad\quad\quad
   \includegraphics[width=3.0in]{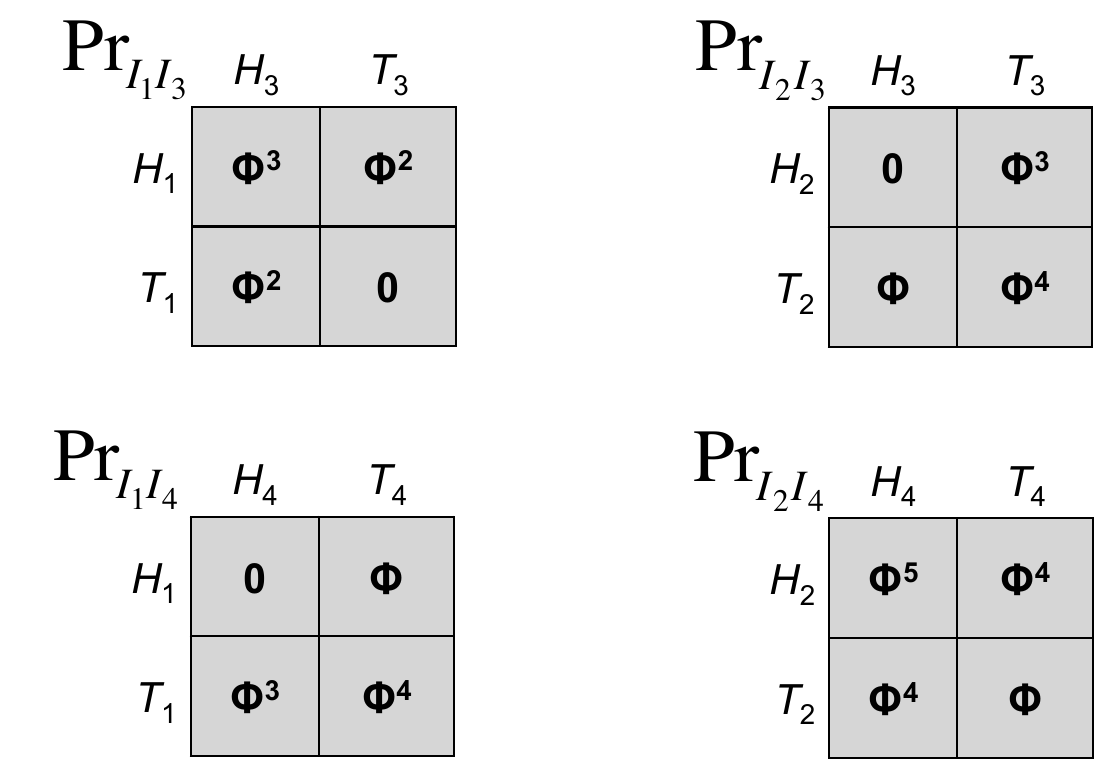}
   \begin{center}
   Figure 8
   \end{center}

\end{figure}

By Proposition 5.2, we know that the signal structure of Figure 8 cannot be realized classically.  (This can also be verified directly; see Appendix C.)  It can, however, be realized quantum mechanically (see Hardy \cite[1993]{hardy93}).  The physical mechanism involves the creation of what is called an entangled pair of particles.  The basic set-up is that two particles --- two photons, for example --- are prepared in a special state and sent off on different trajectories.  Each particle then enters a detector, placed some distance from the source on that particle's trajectory.  Detectors have various settings, and the setting chosen determines which property of a particle is measured.  For example, a detector might be set to measure the so-called spin of a photon along a particular direction.  The outcome of each measurement is binary and can take one of two values, conventionally labelled spin $+1$ or spin $-1$.

Such a quantum system can be used to generate the signal structure of Figure 8.  The spin of one particle is measured at information set $I_1$ or $I_2$.  It is measured along one direction at $I_1$ and along a different direction at $I_2$.  In either case, the measurement has two possible outcomes.  We call them \textit{Heads}$_1$ or \textit{Tails}$_1$, and \textit{Heads}$_2$ or \textit{Tails}$_2$, respectively.  The spin of the second particle is measured at information set $I_3$ or $I_4$.  It is measured along one direction at $I_3$ and along a different direction at $I_4$.  Again, in either case, the measurement has two possible outcomes.  We call them \textit{Heads}$_3$ or \textit{Tails}$_3$, and \textit{Heads}$_4$ or \textit{Tails}$_4$, respectively.  This gives us the form of the signal structure of Figure 8.  The specific probabilities come from the preparation of a particular entangled quantum state (Hardy \cite[1993]{hardy93}, Mermin \cite[1994]{mermin94}).

The choice of a tree with imperfect recall (Figure 7) and of a signal structure with indistinguishability (Figure 8) was quite deliberate.  If a tree has perfect recall, then no signal --- even quantum --- can bring any improvement.  This is not surprising since there is nothing for team members to learn about one another; see Appendix A for a formal argument.  As for indistinguishability, this is a necessary feature of any signal structure that is built using quantum information resources.  This follows from an important property of quantum mechanics called ``no signaling" (Popescu and Rohrlich \cite[1996]{popescu-rohrlich-94}).

\section{Isbell Trees}
We now examine a class of non-Kuhn trees first studied by Isbell \cite[1957]{isbell-57}.  We have already seen an example of an Isbell tree in Figure 1.  This will be the first example where it matters which formulation of classical signals we choose.

We first dispatch the case of perfectly correlated signals.  We argue that any strategy profile using perfectly correlated signals cannot do better than a strategy that does not use any signals.  Indeed, observe that a strategy profile based on perfectly correlated signals cannot prescribe different moves at two nodes in the same information set.  But then, for each realization of the signal, the resulting common move can be replicated as part of a (deterministic) strategy profile that simply prescribes this common move at both nodes.

\begin{figure}[here]
\quad\quad\quad\quad\quad\,\,
   \includegraphics[width=3.7in]{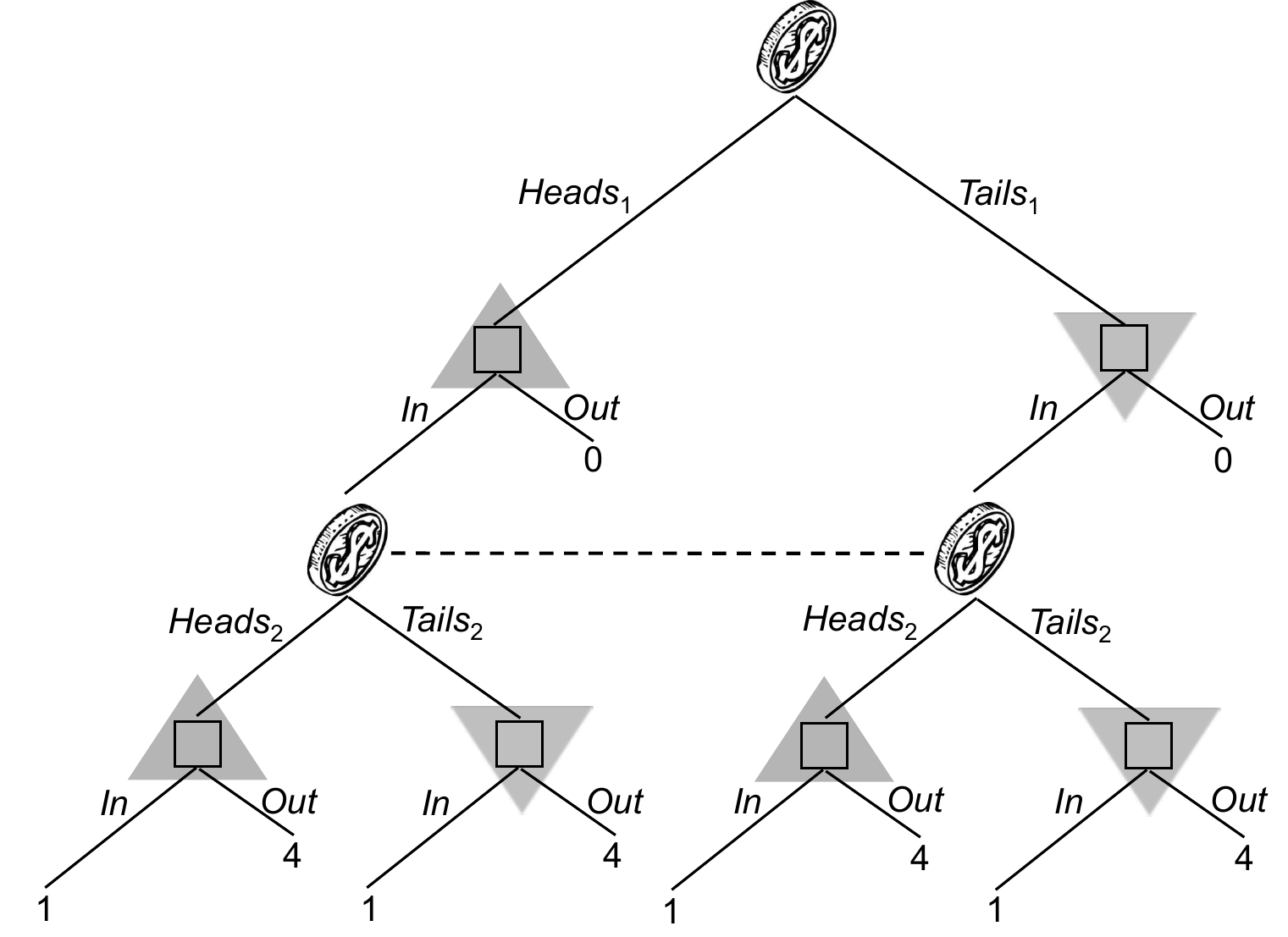}
   \begin{center}
   Figure 9
   \end{center}
\end{figure}

It is no longer true that classical signals have no effect in Isbell trees, once we move from perfectly correlated to i.i.d.~signals.  We review Isbell's \cite[1957]{isbell-57} argument.  We go back to the tree of Figure 1 and extend it by adding two coins at the information set $I$.  The extended tree is depicted in Figure 9.  The coin which is tossed at the root of the tree comes up  \textit{Heads}$_1$ or \textit{Tails}$_1$, and the coin tossed at the subsequent node comes up \textit{Heads}$_2$ or \textit{Tails}$_2$.  The team has two information sets in the extended tree --- one where team members see a coin land \textit{Heads}, and one where they see a coin land \textit{Tails}.  The three nodes in the first information set are shaded with the right-side-up triangles, and the three nodes in the second information set with upside-down triangles.

In this tree, with either no signals or perfectly correlated signals, the team's best expected payoff is $1$ (from choosing \textit{In}).  Now let the signals be i.i.d., as in Figure 10a, and suppose team members adopt the strategy of choosing \textit{In} at the first information set in Figure 9 (right-side-up triangles) and \textit{Out} at the second information set (upside-down triangles).  Then, the team's expected payoff is $1/2 \times 0 + 1/4 \times 4  + 1/4 \times 1 = 5/4 > 1$.  This effect of i.i.d.~signals was first noted by Isbell \cite[1957]{isbell-57}.  Under exchangeability, the team can do even better.  For the signal distribution of Figure 10b, its expected payoff is $1/2 \times 0 + 1/2 \times 4 = 2$.

\begin{figure}[here]
\quad\quad\quad\quad\quad\quad\quad\quad\quad\quad
   \includegraphics[width=2.7in]{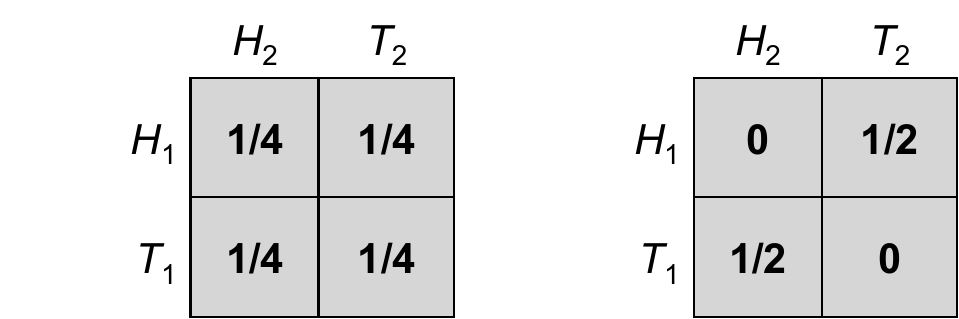}
   \begin{center}
   Figures 10\,a,\,b
   \end{center}
\end{figure}

Cabello and Calsamiglia \cite[2005]{cabello-calsamiglia05} also studied the game of Figure 1 and showed that the availability of quantum signals there allows a team to achieve an expected payoff of $2$.  We have just seen that one does not need to resort to quantum signals in this tree to obtain this improvement.  However, one can easily build other Isbell trees where quantum signals can improve still further on classical signals.  For example, we could simply glue together the tree of Figure 7 (where quantum signals improve on classical signals) and the tree of Figure 1 (where i.i.d.~or exchangeable classical signals improve on no signals).  This would yield an Isbell tree where quantum signals improve on all classical signals.

\section{An Economic Application}
A natural scenario in which team problems arise but direct communication is impossible is high-frequency financial trading.  As a concrete example of communication limitations in this setting, consider two markets located in New York and Shanghai, respectively.  Typically, a new trade is accepted every $0.5$ millisecond by the stock exchange servers.  Even at the speed of light, communication between the two locations takes approximately $40$ milliseconds.  This makes classical arbitrage impossible, since any information about prices on one exchange is already out of date once it reaches the other exchange (Wissner-Gross and Freer \cite[2010]{wissnergross-freer10}).

Now consider a team problem involving two markets (1 and 2) and two traders (Ann and Bob) engaged in local high-frequency strategies.  We assume that the traders are located at a significant distance from each other and from the two markets.  The distances are such that communication prior to their trading decisions is too slow.  We will show how access to quantum signals can enable the two traders to improve their joint performance relative to any classical signals.  The mechanism is based on a well-studied quantum set-up going back to Bell \cite[1964]{bell64} and discussed as a team decision problem in La Mura \cite[2005]{lamura05}.

There are three assets $X$, $Y$,and $Z$, and, at each point in time, each trader needs to sell one of the three assets (chosen with equal probability) against the other two.  When Ann and Bob want to sell the same asset, they do better trading on separate markets, in which case they get payoffs of $0$ (a normalization), rather than on the same market, where they would directly compete against each other and get payoffs of $-M$.  When Ann and Bob want to sell different assets, they do better trading on the same market since each increases the demand for the asset the other wants to sell.  This yields both a payoff of $+m$, as compared with $0$ if they trade on separate markets.  We assume $m < M$.  (This inequality makes sense since even if they sell different assets, they still compete in purchasing the third one.)

If $M$ is sufficiently large compared with $m$, any good pair of strategies for the traders must preclude their selling the same asset on the same market.  In fact, the following is optimal for the traders.  Ann goes to market 2 only when she needs to sell asset $Z$, while Bob goes to market 1 only when he needs to sell $Z$.  (By symmetry, we can replace $Z$ by $X$ or $Y$, and market 2 with market 1.)  To calculate the resulting expected payoff to the traders, note that there are nine equally likely cases according to whether Ann wants to sell asset $X$, $Y$, or $Z$, and similarly for Bob.  In four of these cases, the above strategy profile secures a payoff of $+m$, and otherwise $0$.  So, the expected payoff is $4/9 \times m$.  This is under the assumption of no signals, but, since the scenario corresponds to a Kuhn tree (with imperfect recall), we know from Proposition 5.2 that the addition of classical signals cannot improve the baseline payoff.

Now bring in quantum information resources.  Specifically, we give the traders access to an entangled quantum system on which they can make certain measurements and thereby condition their choices.  Specifically, we assume that the system consists of two particles, one per trader, prepared in the so-called Bell state.\footnote{Bell \cite[1964]{bell64}.  This is a routine set-up in the laboratory.}  This gives rise to the following signal structure:

\begin{figure}[here]
\quad\quad\quad\quad\quad\quad\quad\quad\,
   \includegraphics[width=3.1in]{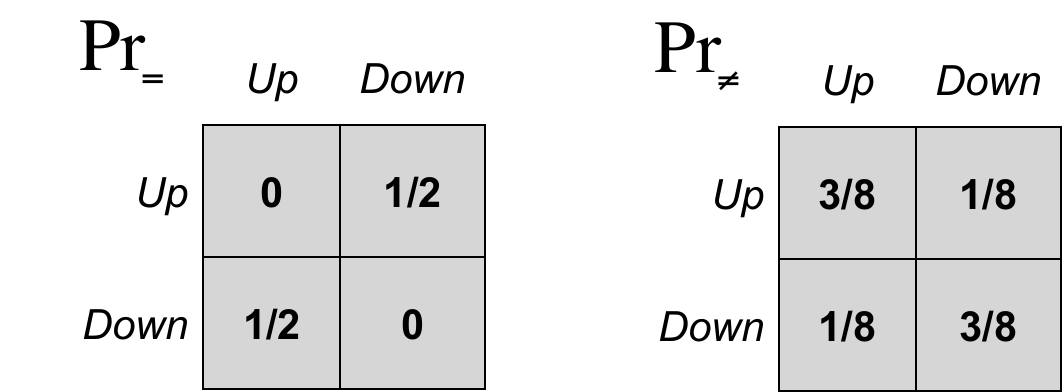}
   \begin{center}
   Figure 11
   \end{center}
\end{figure}

Each trader chooses one of three possible local measurements on the system, which, for convenience, we also label $X$, $Y$, or $Z$.  The left-hand table gives the probabilities of the joint outcomes (each outcome can be \textit{Up} or \textit{Down}) when the traders make the same choice of measurement, and the right-hand table gives the probabilities when they make different choices of measurement.  Consider the following strategy for Ann.  If she wants to sell $X$, then she performs measurement $X$ and, if she observes \textit{Up}, she executes the trade on market 1, while if she observes \textit{Down}, she executes the trade on market 2.  Similarly, if Ann observes $Y$ or $Z$, she performs the corresponding measurement and acts accordingly.  Bob adopts the same strategy.  The expected payoff is calculated as follows.  These strategies always avoid selling the same asset on the same market.  Moreover, in each of the six cases where the traders want to sell different assets, they manage, with probability $3/8 + 3/8 = 3/4$, to trade on the same market.  This leads to an expected payoff of $6/9 \times 3/4 \times m = 1/2 \times m$, which is greater than the baseline payoff of $4/9 \times m$.

\appendixtitleon
\appendixtitletocon
\begin{appendices}

\section{Definition of a Kuhn Tree}
The presentation in this section follows those in Hart \cite[1992]{hart92} and Brandenburger \cite[2007]{brandenburger07}.

\begin{definition}
A (\textbf{finite}) \textbf{Kuhn decision tree} consists of:

\begin{enumerate}
\item[(a)] A set of two players, one called the \textbf{decision maker} (\textbf{DM}) and the other called \textbf{Nature}.

\item[(b)] A finite rooted tree.

\item[(c)] A partition of the set of non-terminal nodes of the tree into two subsets denoted $N$ (with typical element $n$) and $M$ (with typical element 
$m$).  The members of $N$ are called \textbf{decision nodes}, and the members of $M$ are called \textbf{chance nodes}.

\item[(d)] A partition of $N$ (resp. $M$) into \textbf{information sets} denoted $I$ (resp. $J$) such that for each $I$ (resp. $J$):

\begin{enumerate}
\item[(i)] all nodes in $I$ (resp. $J$) have the same number of outgoing branches, and there is a given 1-1 correspondence between the sets of outgoing branches of different nodes in $I$ (resp.$J$);

\item[(ii)] every path in the tree from the root to a terminal node crosses each $I$ (resp. $J$) at most once.
\end{enumerate}
\end{enumerate}
\end{definition}

For each information set $I$ (resp. $J$), number the branches going out of each node in $I$ (resp. $J$) from $1$ through \#$I$ (resp. \#$J$) so that the 1-1 correspondence in (d.i) above is preserved.

\begin{definition}
A \textbf{strategy} (for the DM) associates with each information set $I$, an integer between $1$ and \#$I$, to be called the DM's \textbf{choice} at $I$.  Let $S$ denote the set of strategies for the DM. \ A \textbf{state of the world} (or \textbf{state}) associates with each information set $J$, an integer between $1$ and \#$J$ to be called the \textbf{choice} of Nature at $J$.  Let $\Psi$ denote the set of states.
\end{definition}

Note that a pair $(s,\psi)$ in $S \times \Psi$ induces a unique path through the tree.

\begin{definition}
Fix a node $n$ in $N$ and a strategy $s$.  Say $n$ is \textbf{allowed under} $s$ if there is a state $\psi$ such that the path induced by $(s,\psi)$ passes through $n$.  Say an information set $I$ is \textbf{allowed under} $s$ if some $n$ in $I$ is allowed under $s$.
\end{definition}

\begin{definition}
Say the DM has \textbf{perfect recall} if for any strategy $s$, information set $I$, and nodes $n$ and $n^{\ast}$ in $I$, node $n$ is
allowed under $s$ if and only if node $n^{\ast}$ is allowed under $s$.
\end{definition}

\begin{definition}
Say a node $n$ in $N$ is \textbf{non-trivial} if it has at least two outgoing branches.
\end{definition}

Define a relation of precedence on the DM's information sets $I$, as follows: Given two information sets $I$ and $I^{\prime}$, say that $I$ \textbf{precedes} $I^{\prime }$ if there are nodes $n$ in $I$ and $n^{\prime}$ in $I^{\prime}$ such that the path from the root to $n^{\prime}$ passes through $n$.  It is well known that if the DM has perfect recall and all decision nodes are non-trivial, then this relation is irreflexive and transitive, and each information set $I$ has at most one immediate predecessor.  (Proofs of these assertions can be found in Brandenburger \cite[2007, Appendix]{brandenburger07}, or can be constructed from arguments in Wilson \cite[1972]{wilson-72}.)

Kuhn \cite[1953, p.213]{kuhn-53} observes that perfect recall implies that the DM remembers: (i) all of his choices at previous information sets; and (ii) everything he knew at those information sets.  The following two lemmas formalize these observations.  (The proofs are in \cite[2007, Appendix]{brandenburger07}.)

\begin{lemma}
Suppose the DM has perfect recall and all decision nodes are non-trivial.  Fix information sets $I$ and $I^{\prime}$, and strategies $s$ and $s^{\prime}$.  Suppose that $I^{\prime}$ is allowed under both $s$ and $s^{\prime}$, and $I$ precedes $I^{\prime}$.  Then $I$ is allowed under both $s$ and $s^{\prime}$, and $s$ and $s^{\prime}$ coincide at $I$.
\end{lemma}

Next, write:
\begin{equation*}
\begin{array}{c}
[I] = \{ \psi : I \,\text{is allowed under}\,\, \psi \}.
\end{array}
\end{equation*}

\begin{lemma}
Suppose the DM has perfect recall and all decision nodes are non-trivial.  Fix information sets $I$ and $I^{\prime}$.  If $I^{\prime}$ succeeds $I$, then $[I^{\prime}] \subseteq [I]$.
\end{lemma}

We can now justify the entries in the first row of Table 1 in the text.  Fix a perfect-recall tree and an information set in the tree.  Given that the DM remembers all of his choices at previous information sets, and everything he knew at those information sets, the only effect of a signal (classical or quantum) at the information set can be to change the DM's probabilities about the state of the world.  But, we have assumed (see Section 5 in the text and Appendix B below) independence between Nature and signals, so no such change is possible.

\section{Correlation Between Nature and Signals}
Consider the imperfect-recall decision tree depicted in Figure B.1, and the joint distribution of a coin toss at the team's information set, and Nature's move, depicted in Figure B.2.  The team gets an expected payoff of $1$ (in fact, $1$ almost surely) by choosing \textit{Left} after \textit{Heads} and \textit{Right} after \textit{Tails}.  Its expected payoff in the underlying tree is $1/2$ (regardless of his strategy).

\begin{figure}[here]
\quad\quad\quad\quad\quad\quad\quad\quad\quad\quad\quad\quad\quad\quad
   \includegraphics[width=2.1in]{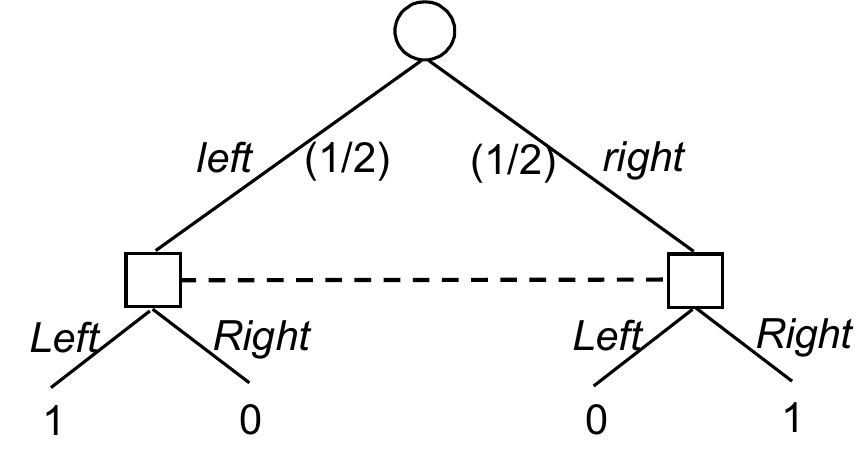}
   \begin{center}
   Figure B.1
   \end{center}
\end{figure}

\begin{figure}[here]
\quad\quad\quad\quad\quad\quad\quad\quad\quad\quad\quad\quad\quad\quad\quad\,\,
   \includegraphics[width=1.2in]{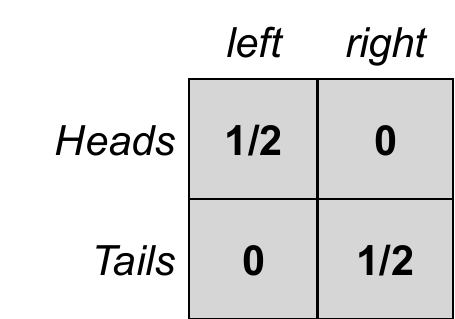}
   \begin{center}
   Figure B.2
   \end{center}
\end{figure}

This shows that Proposition 5.2 is false in the absence of the independence assumption between Nature and signals.

\section{Non-Existence of a Joint Probability Space}
It is instructive to see directly (rather than via appeal to Proposition 5.2) why the signal structure of Figure 8 cannot be derived from a joint probability space $(\Omega, \mu)$.  We can take $\Omega$ to be as depicted in Table C.1,\footnote{There is no loss of generality in taking $\Omega$ to be of this canonical form.  This follows from starting with a general (even infinite) probability space and taking the image measure.} so that $\mu$ is a probability measure on the set $\{\omega^0, \ldots, \omega^{15}\}$.  Now let's try to derive the signal structure of Figure 8.  The conditions ${\rm{Pr}}_{I_{1}I_{3}}({\rm\textit{Tails}}_1, {\rm\textit{Tails}}_3) = 0$, ${\rm{Pr}}_{I_{2}I_{3}}({\rm\textit{Heads}}_2, {\rm\textit{Heads}}_3) = 0$, and ${\rm{Pr}}_{I_{1}I_{4}}({\rm\textit{Heads}}_1, {\rm\textit{Heads}}_4) = 0$ require, respectively:
\begin{eqnarray*}
\mu(\omega^{10}) + \mu(\omega^{11}) + \mu(\omega^{14}) + \mu(\omega^{15}) & = & 0, \\
\mu(\omega^0) + \mu(\omega^1) + \mu(\omega^8) + \mu(\omega^9) & = & 0, \\
\mu(\omega^0) + \mu(\omega^2) + \mu(\omega^4) + \mu(\omega^6) & = & 0,
\end{eqnarray*}

\noindent while the condition ${\rm{Pr}}_{I_{2}I_{4}}({\rm\textit{Heads}}_2, {\rm\textit{Heads}}_4) = \Phi^5 > 0$ requires:
\begin{equation*}
\mu(\omega^0) + \mu(\omega^2) + \mu(\omega^8) + \mu(\omega^{10}) > 0,
\end{equation*}

\noindent which is then impossible.

\begin{figure}[here]
\quad\quad\quad\quad\quad\quad\quad\quad\quad\quad\quad\quad\quad\quad\,
   \includegraphics[width=2.0in]{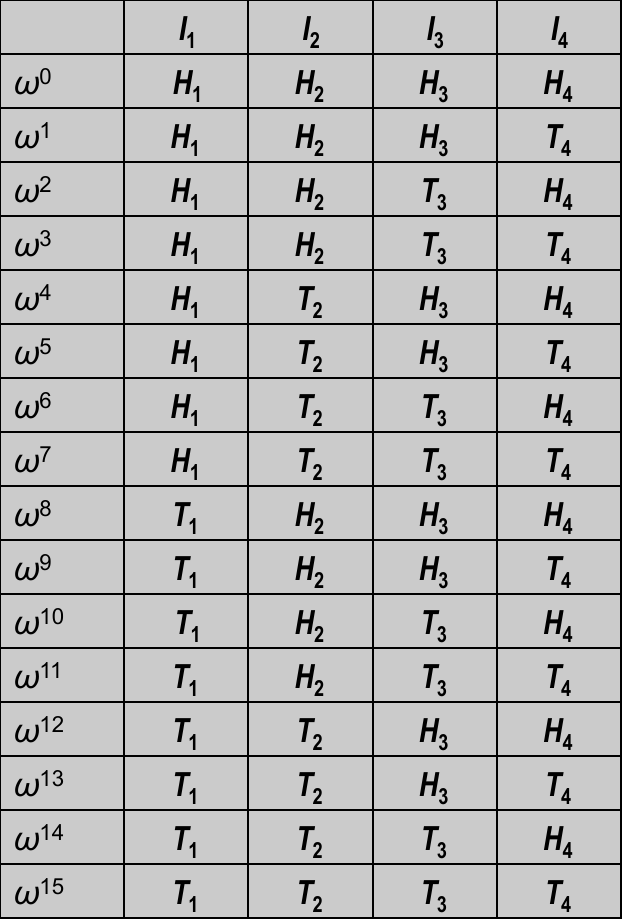}
   \begin{center}
   Table C.1
   \end{center}
\end{figure}

In quantum mechanics this impossibility argument is usually described as a strengthened version of the famous Bell's Theorem (Bell \cite[1964]{bell64}).  (The strengthening is precisely the point of Hardy \cite[1993]{hardy93}.)  The lesson of Bell's Theorem is that quantum mechanics allows \textbf{non-local correlation}, that is, dependence among particles that cannot be viewed as arising from common-cause correlation.  The common cause, if it existed, would arise from the operation of extra variables (usually called \textbf{hidden variables}) equipped with a classical probability distribution.  Since quantum mechanics allows correlation that is not common-cause, and since particles can be physically distant from each other at the point of measurement, the correlation is called non-local.

Non-existence of a joint probability space can be understood as arising from the \textbf{incompatibility} of certain measurements on quantum systems.  Measurements of position and momentum make the most famous example.  In the physical set-up underlying Figure 8, there are two incompatibilities --- one between the two different measurements on the first particle, and another between the two different measurements on the second particle.

\end{appendices}

\end{document}